\documentclass[sigconf]{acmart}
\usepackage{setspace}
\usepackage{censor}
\usepackage{balance}
\AtBeginDocument{%
  }

\acmYear{2025}
\acmDOI{XXXXXXX.XXXXXXX}
\acmConference[LIMITS '25]{11th Workshop on Computing within Limits}{June 26 --27, 2025}{Online} 

\setcopyright{none}
\acmDOI{10.48550/arXiv.2508.05992}
\acmISBN{}

\begin{document}

\title{Surviving the Narrative Collapse: Sustainability and Justice in Computing Within Limits}


\author{Dave Guruge}
\affiliation{%
  \institution{College of Work-Based Learning, Otago Polytechnic}
  \city{Tauranga}
  \country{New Zealand}
}
\email{dguruge@me.com}

\author{Samuel Mann}
\affiliation{%
  \institution{College of Work-Based Learning, Otago Polytechnic}
  \city{Dunedin}
  \country{New Zealand}
}
\email{Samuel.Mann@op.ac.nz}

\author{Ruth Myers}
\affiliation{%
  \institution{College of Work-Based Learning, Otago Polytechnic}
  \city{Dunedin}
  \country{New Zealand}
}
\email{Ruth.Myers@op.ac.nz}

\author{Oliver Bates}
\affiliation{%
  \institution{Lancaster University}
  \city{Lancaster}
  \country{United Kingdom}
}
\email{o.bates@lancaster.ac.uk} 

\author{Mikey Goldweber}
\affiliation{%
  \institution{Denison University}
  \city{Ohio}
  \country{United States}
}
\email{goldweberm@denison.edu} 

\author{Andy Williamson}
\affiliation{%
  \institution{Democratise}
  \city{Isle of Skye}
  \country{Scotland}}
\email{andy@democrati.se} 

\author{Jon Lasenby}
\affiliation{%
  \institution{Maranga Ltd}
  \city{Nelson}
  \country{New Zealand}
}
\email{jonolasenby@gmail.com} 

\author{Ian Brooks}
\affiliation{%
  \institution{University of the West of England}
  \city{Bristol}
  \country{United Kingdom}
}
\email{ian.brooks@uwe.ac.uk} 
\renewcommand{\shortauthors}{Guruge et al.}

\begin{abstract}
  Sustainability-driven computing research—encompassing equity, diversity, climate change, and social justice—is increasingly dismissed as 'woke' or even dangerous in many sociopolitical contexts. As misinformation, ideological polarisation, deliberate ignorance and reactionary narratives gain ground, how can sustainability research in computing continue to exist and make an impact? This paper explores these tensions through Fictomorphosis, a creative story retelling method that reframes contested topics through different genres and perspectives. By engaging computing researchers in structured narrative transformations, we investigate how sustainability-oriented computing research is perceived, contested, and can adapt in a post-truth world.
\end{abstract}

\keywords{taboo, post-truth, post-qualitative, methodological, story-telling }

\maketitle

\section{Introduction}

\noindent
\begin{center}

In this paper,\\
we start not at the beginning,\\
but at the edge of the unsayable,\\
where silence is not absence,\\
but something done to us.
 
\end{center}
\vspace{6pt}

For many years, computing research has largely ignored, or perhaps tolerated, the small band of climate change, sustainability, and social justice (and so on) activists within its ranks.  Now they are finding their work increasingly marginalised. Labelled as “woke,” “dangerous,” or politically inconvenient, these research areas face funding cuts, institutional pushback, and ideological dismissal. This paper introduces a method designed to provide insight into how responsible computing researchers experience and might respond to such conditions where their research is considered taboo.  Rather than retreating, we open the dialogue for reframing practices, forming alliances, and finding alternative ways to sustain this suppressed work. 

We describe a constrained-creative story-telling approach that has allowed for an exploration of the consequences of systematic knowledge suppression through a diverse array of literary forms. From children’s stories to academic papers, police reports to poetry, each piece examines what happens when words and ideas are first hidden then forbidden. Crucial information disappears behind redacted black bars, obscuring meaning and revealing how censorship creates dangerous voids.

Recent developments have seen significant reductions in research funding globally, particularly in areas intersecting with sustainability, climate change, and social justice.  US Senator Ted Cruz released a database of over 3,400 research grants he deemed to promote a "far-left ideology," targeting research areas including social justice, gender, race, and environmental justice \cite{insidehighered2025}. This list includes projects focused on computational decision-making for climate resilience and community-based technological innovation. It further places at risk applied computing initiatives that confront systemic racism in mathematics education, expand equitable access to technological infrastructure, strengthen food systems, and advance racial and gender equity in STEM fields. The National Science Foundation has subsequently terminated over 380 grants totalling approximately \$233 million, affecting projects related to diversity, equity, inclusion (DEI) \cite{science2025} \cite{ft2025}, and misinformation \cite{apnews2025}.  The Trump administration has also proposed significant budget cuts to NASA and NOAA, threatening climate science and environmental monitoring programmes \cite{guardian2025}. This issue is not restricted to the US.  In New Zealand, for example, the government restructured the Marsden Fund, redirecting focus toward "core sciences" and effectively ending funding for humanities and social sciences, impacting interdisciplinary research, including computer science projects addressing societal challenges \cite{nzherald2024}.  In the UK, a restructuring of the Turing Institute and the cancellation of the Global Challenges Research Fund have disproportionately affected environmental and social science, particularly that concerning the Global South \cite{nwako2023doing}. All this leads to questions of whether even the Sustainable Development Goals - labelled `The Enemy'' to techno-optimism \cite{andreessen2023techno} - can survive? \cite{grove2025usfundingcuts}. Not just data \cite{bbc2025climatedata}, but whole areas of science are being expunged. 

Such deliberate ignorance is not new \cite{proctor2008agnotology}. Organised ignorance is a systemic phenomenon in organisations where knowledge is deliberately obscured, avoided, or left unexamined to maintain existing power structures and social norms. It operates as a protective mechanism that upholds taboos by preventing the disclosure or discussion of sensitive issues, thereby sustaining institutional silence. Organised ignorance plays a crucial role in shaping discourse around sensitive technology-related topics such as censorship, surveillance, and ethical hacking. Organisations and governments strategically manage ignorance by suppressing conversations that challenge the status quo, ensuring that critical debates remain on the fringes. 

Organised ignorance is not merely a passive void but a structured and strategic phenomenon that benefits those in power by limiting awareness and accountability. ``Agnotology'' the study of ignorance, reveals a history and political geography \cite{proctor2008agnotology}. In the realm of technology, this manifests through deliberate ambiguity around digital rights, encryption policies, and government surveillance programs, ensuring that individuals remain uncertain or uninformed about the extent of control exercised over online spaces \cite{mcgoey2012logic}. Collective ignorance is sustained within organisations through unspoken agreements that certain topics are too controversial or disruptive to address openly \cite{bakken2023organised}. 
\section{Literature}
“Taboo” topics in computing refer to subjects considered off-limits or too sensitive to discuss openly within the tech community or society. In general, a taboo denotes something forbidden in a culture, upheld less by laws than by social disapproval \cite{champion2023taboo, kannabiran2011sexuality}. In a computing context, this category includes ethical issues, practices, or research areas that provoke discomfort, controversy, or even censorship due to perceived moral, legal, or security transgressions. In other words, these are computing topics that people avoid or tiptoe around because they challenge prevailing norms or power structures \cite{kannabiran2011sexuality}.

Several ethical, industry, and societal factors can render a computing topic taboo. Modern technologies often outpace existing norms and regulations, creating a “policy vacuum” that leads to uncertainty and debate \cite{kannabiran2011sexuality}. When there are no established policies for a new computer capability, society can struggle to decide how it should be used. Topics that threaten powerful interests (for example, state surveillance programmes or a company’s reputation) may be suppressed or avoided in public discourse \cite{moor1985ethics}. Societal values and fears also play a role: if a tech practice appears to violate core values like privacy, fairness, or safety, it may be stigmatised and met with public backlash \cite{fontes2022ai}. In sum, contentious tech issues often become taboo when they highlight unresolved ethical questions or conflicts between stakeholder interests.

There are existing examples of taboo areas in computing. Sexuality research in HCI faces censorship, moral judgment, and institutional resistance, forcing scholars to reframe studies (e.g., as "intimacy" instead of "sexuality") to gain approval, limiting discourse and innovation \cite{kannabiran2011sexuality}. Algorithmic bias, where algorithms reflect human prejudices, remains a sensitive subject, as shown when Amazon scrapped a résumé screening tool that discriminated against women \cite{starke2022fairness,bayana2024bias}. Addressing bias is crucial yet politically uncomfortable, leading to underreporting and avoidance.

Mass digital surveillance, especially in contexts like China, pits public safety against privacy, with governments and industries downplaying ethical debates to avoid scrutiny \cite{lu2005privacy,adar2013benevolent,bauman2014after,lydon2025community}. Discussions about censorship are similarly constrained: platforms and governments often justify secretive moderation practices like shadow banning, yet this opacity limits user rights and chills debate about free expression \cite{clark2017shifting,leerssen2023shadow}.

Ethical hacking also occupies a grey area, where teaching or practicing offensive security techniques raises legal and moral concerns \cite{jaya2024ethics,saha2020ethical}. Rather than being fully embraced as a cybersecurity necessity, hacking topics remain fraught, reinforcing controlled ignorance as a strategy for maintaining authority, echoing Foucault’s critique of knowledge control \cite{gross2015ignorance}.
This controlled ignorance ensures that critical reflections on digital ethics remain suppressed, preventing a broader societal reckoning with the implications of mass surveillance, internet restrictions, and cybersecurity policies.

The constriction of permissible discourse can be analysed through the framework of the Overton window, which delineates the spectrum of ideas deemed politically or socially acceptable at any given moment. Overton originally articulated the concept of a Window of Acceptability, highlighting how specific ideas transition from being deemed unthinkable or fringe to becoming part of mainstream discourse as social norms evolve \cite{mackinac2019overton, krick2023buyin}. As certain fields of computing research — such as climate justice or algorithmic equity — transition outside this window, they face stigmatisation, defunding, or become unspeakable, thereby reinforcing their taboo status not through evidence, but due to evolving ideological norms.

In this paper, we attempt to shift the Overton window by reasserting the legitimacy of these marginalised research areas and demonstrating how they are essential to a just, sustainable, and critically reflective computing discipline. We first introduce a new method "Fictomorphosis", then use this approach to write stories.  In the last section we explore these stories and reflect on the collaborative process. 

\section{Method: Fictomorphosis}
This paper offers a novel methodological contribution for addressing taboo topics particularly where insider researchers use ethnography and autoethnography to explore deeper meanings and generate insights that drive positive change. Insider researchers, who often have privileged access to organisations, communities, and lived experiences, rely on compelling narratives to motivate change-based research. However, some events and contexts remain off-limits due to non-disclosure agreements (NDAs), privacy concerns, safety risks, ethical constraints, the stigma surrounding the topic and the “taboo” nature of the topic. These barriers create significant challenges for emancipatory research, where issues like misogyny, racism, workplace injustice, and other forms of oppression remain hidden or unexplored as “taboo” topics.

Despite the importance of bringing these issues to light, researchers lack effective methodologies for engaging with sensitive, restricted, or ethically complex subjects. Traditional ethnographic and autoethnographic approaches attempt to mitigate ethical risks by anonymising participants, settings, and events. However, this raises concerns about the extent of fabrication involved, leading to questions about authenticity, representation, and research ethics. Moreover, anonymisation does not always fully protect individuals, particularly in small or insular communities where indirect identification remains possible.

In response to these challenges, we use “Fictomorphosis”, a creative and ethical research method designed to engage with taboo topics while protecting individuals and transforming real-world experiences into new, fictionalised narratives which are subsequently used to derive meaningful insights, challenge dominant narratives, and foster critical reflection. This is the first description of Fictomorphosis, for a more comprehensive discussion, see \cite{guruge2025fictomorphosis, guruge2025DPP}. 
Fictomorphosis is a narrative transformation method that allows researchers to engage with restricted or taboo topics by decontextualising and reimagining them through creative storytelling. Rather than anonymising participants or obscuring details, Fictomorphosis embraces fictionalisation to distance the narrative from its origins while preserving the essence of the lived experience.

The process involves reducing an experience, insight, or ethical dilemma to a single core question or statement (the “nugget”) and then retelling it in an entirely new genre or context — such as science fiction, fantasy, poetry, allegory, or fable. These fictionalised accounts become the foundation for thematic analysis and practice-based change.
A key benchmark of Fictomorphosis is that original participants in the research environment would see the stories as perceptive, yet they would not recognise the stories' origin in their own experience. This ensures that while the emotional truth and ethical complexity of the original experience remain intact, the identities, settings, and specificities are fully transformed.

\section{Methodology: How Fictomorphosis Works}
Fictomorphosis is a structured, iterative process that consists of five key stages:
\begin{itemize}
  \item Extracting the Nugget
  \begin{itemize}
    \item The researcher refines an experience or insight into a core question or statement that encapsulates its ethical, social, or emotional essence, following a process of reflective resonance writing and provocation development grounded in the diffractive method \cite{Barad2014}.
    \item For instance, in a study on algorithmic bias in hiring, the nugget might be: “What happens when a system designed for fairness inherits the biases of the past?”
  \end{itemize}
  \item Model Interrogation
  \begin{itemize}
    \item The nugget is subjected to a self-questioning process, drawing from philosophical, ethical, and critical perspectives.
    \item This interrogation helps uncover hidden assumptions, systemic forces, and contextual nuances that shape the experience.
  \end{itemize}
  \item Genre Transformation
  \begin{itemize}
    \item The researcher retells the core narrative in multiple genres, such as:
    \begin{itemize}
      \item Science fiction (exploring bias through an AI dystopia)
      \item Fable (recasting power hierarchies in an animal kingdom)
      \item Poetry (distilling emotional weight into a symbolic form)
    \end{itemize}
    \item These transformations allow researchers to shift perspectives, reveal underlying structures, and engage broader audiences in ways that traditional research narratives may not.
  \end{itemize}
  \item Ethical Anonymisation Check
  \begin{itemize}
    \item This stage ensures that all elements of the narrative are sufficiently fictionalised, protecting identities while preserving the ethical and thematic core.
    \item This step is especially critical in research on marginalised communities, workplace oppression, or cybersecurity, where real-world disclosure might have harmful consequences.
  \end{itemize}
  \item Crystallisation and Diffraction
  \begin{itemize}
    \item Instead of seeking a single “truth,” Fictomorphosis uses crystallisation—a method of examining an issue through multiple, intersecting perspectives to extract potentials and construct meaningful tools
    \item Diffraction techniques help break down dominant narratives and reconstruct them to challenge power structures and assumptions.
      \item From the transformed narratives, the researchers identify emerging concepts, patterns, and tensions that stand out—these are the potentials within the stories.
    \item These potentials are then synthesised into meaningful models, conceptual frameworks, or practical strategies that contribute to professional practice, practitioner development, and wider academic discourse.
    \item For instance, in the case of algorithmic bias, insights drawn from different genres might reveal new ethical guidelines for algorithmic governance, alternative design paradigms, or critical frameworks for evaluating fairness in automated systems.
    \item This final stage ensures that the research does not remain at the level of creative exploration but is translated into tangible contributions that inform and shape both theory and practice.
  \end{itemize}
\end{itemize}
\subsection{Application of Fictomorphosis in Professional Practice}
Fictomorphosis is particularly valuable for Professional Practice researchers \cite{guruge2025fictomorphosis}, offering a way to engage with stories that must be told yet cannot be told directly. This method has been tested and refined through a series of experimental applications, demonstrating its ability to address both practical and ethical challenges \cite{guruge2025DPP}. One key application involved research with intellectually disabled adults, a group often excluded from research due to consent-related concerns. By using Fictomorphosis, researchers were able to explore critical themes in professional caregiving, autonomy, and inclusion without compromising ethical boundaries. 

In Fictomorphosis, identifying the nugget entails condensing a genuine experience, ethical dilemma, or revelation into a singular, impactful question or remark that encapsulates the knowledge of the event including its emotional or social core. This approach usually entails a thorough reflective, or more accurately, diffractive journey wherein the author initially composes a short reflective autoethnographic work grounded on personal experience, thereafter engaging in self-interrogation by examining prior literature on the topic. The nugget arises as a concise recommendation from this comprehensive thought and analysis.

\subsection{FictoLimits} 

We applied Fictomorphosis to the question of the suppression of sustainability research within computing. We dubbed this process "Fictolimits". 

\begin{figure}[h]
  \centering
  \includegraphics[width=\linewidth]{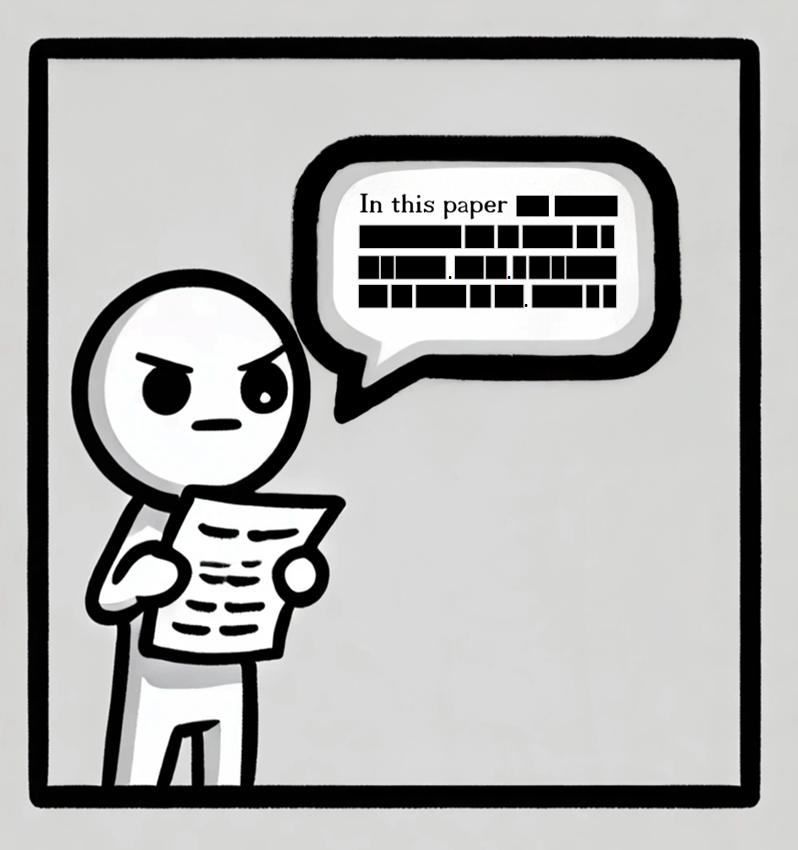}
  \caption{Agreed "nugget" story prompt 
  }
  \Description{A cartoon sick figure with a furrowed brow reads from a piece of paper. A speech bubble reads "In this paper" but the rest is redacted by black bars covering the text.}
  \label{fig:nugget}
\end{figure}

The authors have all previously published in computer science and what might broadly be considered sustainability and computing research. We met online to discuss the idea of using Fictomorphosis to explore the notion of our research being taboo.  This is a slight departure from previous uses of Fictomorphosis that focused on overcoming existing taboos.  After some discussion, we decided to have a nugget that both described the problem, but also demonstrated it in a humorous manner (Figure \ref{fig:nugget}). 

\section{Stories}
The collection of 41 stories \cite{guruge2025redactedBook} explores the consequences of systematic knowledge suppression through a diverse array of literary forms. From children’s stories to academic papers, police reports to poetry, each piece examines what happens when words and ideas are first hidden then forbidden. Crucial information disappears behind redacted black bars, obscuring meaning and revealing how censorship creates dangerous voids. 

Characters discover forbidden desks, face absurd academic committees and struggle to communicate in increasingly constrained languages, while vague threats of unseen disasters loom on the horizon. Misinformation, confusion and invented narratives fill empty spaces; and truth is collateral damage. 
What at first seem fragmented narratives, when collected together, send a cohesive, chilling warning about the risks of controlling discourse and of politicising science, particularly around contentious issues such as climate change and the ethics of the digital world. The systematic erasure of knowledge, as the collection ultimately shows us, deprives society of not only words but of the ability to confront the crises those words describe. What emerges is a world where:
\begin{enumerate}
\item Knowledge and truth are contested
\item Power dynamics control knowledge creation
\item There is institutional capture of narratives
\item We have a meta-awareness of narrative construction
\item There is a fragmentation of meaning
\end {enumerate}

\begingroup
\section*{A poem}
\begin{center}

In the beginning there was [REDACTED]

"you’ve got a week”

his client told him

“I want light and firmament”

surely, he meant filament

God created light
\vspace{10pt}

the client wanted dry land

and land that’s not so dry

God called this “the sea”

making plants for the dry parts
\vspace{10pt}

he added animals 

because they were fun

“no dinosaurs”

the client was adamant 

but he wanted people
\vspace{10pt}

against his judgement

God made them 

he did this last of all

knowing from experience  

things could go awry	

\vspace{10pt}
what he forgot to do

what the client didn’t ask for 

what would have saved the day

was [REDACTED]
\end{center}
\endgroup
\vspace{20pt}
\newpage

\onecolumn

\begingroup
\setstretch{1.1}
\section*{Chapter Something-Or-Other, In Which Words Disappear}

One morning in the Hundred Acre Wood, Winnie-the-Pooh found himself in quite a bother. He had been reading—well, not quite reading, because most of the words in the book Rabbit had given him were covered in big, dark smudges.

“Oh, bother,” said Pooh, tilting his head. “It seems this story is missing all the words.”

Piglet, who was always rather small and sometimes rather worried, peeked over Pooh’s shoulder. “What does it say, Pooh?”

Pooh squinted. “It says… ‘Those words are forbidden’—but, of course, they’re \censor{[REDACTED]}. And oh, Piglet, it is very frustrating.”

Piglet nodded solemnly. “I suppose it’s hard to know what’s not allowed if you don’t know what it was in the first place.”

“Exactly,” Pooh agreed. “But it’s even harder to know what is allowed when you don’t know anything at all.”

At that moment, Rabbit bustled past, muttering something about “woke nonsense” and waving his paws about. “It’s all very simple,” he said importantly. “We don’t need to know. Too much knowing just causes trouble.”

Eeyore, who had been standing nearby in his usual gloomy fashion, flicked his tail. “Mmm. Meanwhile, nonsense spreads faster than a thistle fire.” He sighed. “But I suppose no one listens to Eeyore.”

Pooh and Piglet turned back to the book. The more they stared, the more frustrating the blacked-out words became.

“I think,” said Pooh, after a thoughtful moment, “that when you erase words, people don’t just forget them. They start guessing what they were. And sometimes, they guess all wrong.”

Piglet shivered. “Oh dear. What if they make up something even worse?”

Pooh nodded. “That is the trouble with redacting, Piglet. It makes a bear wonder… if the words were so very important, why take them away at all?”

\vfill
\section*{The Forbidden Desk}
There was a desk in the corner of the old public library in Kandy, tucked away in the deepest shadows. No one used it. No one even spoke about it. It sat there, collecting dust, its once-polished wood dulled by time and silence.

Eliza had worked at the library long enough to notice strange things. Books would vanish without a trace, entire sections of knowledge simply… missing. People would ask about certain topics—history, science, politics—and be met with blank stares or hurried whispers. But it was the desk that intrigued her the most.

There was a plaque on it, or at least, there had been. Now, it was just a tarnished brass plate with deep black scratches. She could almost make out the original inscription beneath the crude, clumsy redactions:

"Those words are forbidden."

But, of course, they were [REDACTED].

Eliza ran her fingers over the deep gouges in the wood. Someone had tried very hard to erase whatever had once been written here. That made her wonder—what was so important that it had to be wiped away entirely?

She asked the head librarian, Mr. Graves, about it. He barely looked up from his newspaper. “Nothing to worry about. Just woke nonsense.”

That was the answer to everything these days.

But Eliza knew better. She had seen how misinformation spread, filling the gaps where real knowledge had been erased. People made up their own truths—fantastical, terrifying stories, whispered between the shelves.

She leaned closer to the desk, studying it. The drawers had been sealed shut, but she could see the faint outline of where they had once opened. With a glance over her shoulder, she pried at the edge of one. The wood creaked, and then—snap—the old lock gave way.

Inside, she found pages. Not books—those could be tracked, confiscated—but loose pages, hidden away. She held one up to the dim light.

Every word had been blacked out.

Every single word.

Her hands trembled. She imagined it all: history erased, stories untold, jobs vanishing, oppression tightening like a vice. Redacted = Frustration.

And yet, as she stared at the blacked-out pages, something strange happened.

Her mind started filling in the blanks.

People don’t forget, she realised. When you take away the truth, they don’t just let it go—they invent. They piece together whatever scraps remain and create something new. And sometimes… sometimes what they create is far more dangerous than what was taken.

In her mind, she saw it all—every lost word, every deleted fact, every vanishing voice. And in the background? A blurred-out building, barely visible—except for the flames licking at its edges.

Yes. It was definitely on fire.

She stuffed the papers back into the desk and pushed the drawer shut. The truth had been hidden, but it wasn’t gone. Not yet.

\endgroup
\newpage

\section*{The Committee for Research Excellence}
\begingroup
\setstretch{1.2}
\hangindent=2em
\hangafter=0

\noindent
(A dim university boardroom. Five stern white men in suits face a nervous researcher, DR MENDEL. A sign on the wall reads: “INQUIRY COMMITTEE: Ensuring Safe, Neutral, and Unthinking Research Since Tuesday.”)

\noindent
CHAIRMAN: Right then, Dr Mendel. You’re here on suspicion of dangerous research.

\noindent
DR MENDEL: What?! I haven’t done anything! 

\noindent
COMMITTEE MEMBER 1: Exactly. Suspiciously silent.

\noindent
COMMITTEE MEMBER 2: We know what you were about to say…

\noindent
COMMITTEE MEMBER 1: Oh, don’t try to deny it! We can tell.
\noindent
DR MENDEL: Tell what? 

\noindent
COMMITTEE MEMBER 2: What you’re thinking of saying. 

\noindent
DR MENDEL: I haven’t said anything yet.

\noindent
CHAIRMAN: Exactly. And that’s the most suspicious thing of all! (slams table) If you had nothing to hide, you’d already be talking about something safe!

\noindent
DR MENDEL: But I don’t even know what you’re accusing me of! 

\noindent
COMMITTEE MEMBER 3: Oh, come on, Mendel. You were about to say the phrase. 

\noindent
DR MENDEL: What phrase?

\noindent
COMMITTEE MEMBER 1: The phrase that starts every piece of woke academic subversion!

\noindent
COMMITTEE MEMBER 2:  (Slowly, almost in a whisper) “In this paper…”
(*Gasps all around. A woman faints. Someone fans her with a large report titled "Approved Research Topics: A Short List (Vol. 1 of 1, Page 1 of 1)".)

\noindent
DR MENDEL: Oh, for the love of—look, I wasn’t going to say anything radical! It’s just an analysis of computing’s impact on—

\noindent
CHAIRMAN: STOP! (points dramatically) We know what you were thinking!

\noindent
DR MENDEL: No, you don’t!

\noindent
COMMITTEE MEMBER 1: Oh yes we do! First, you say, “In this paper…”—that’s how they get you. Then you start talking about climate change, sustainability, social justice… Next thing you know, you’re decolonising algorithms and questioning whether AI should have ethics! (Another collective gasp. Someone bursts into tears. The Chairman hands them a pamphlet titled "Emotional Regulation in the Age of Unquestioned Progress.")

\noindent
DR MENDEL: That’s absurd! I haven’t even written the paper yet! 

\noindent
COMMITTEE MEMBER 3: Exactly! Pre-emptive censorship is the best kind—no messy words to redact!
(They all nod in agreement.)

\noindent
DR MENDEL: You can’t punish me for what I haven’t written! 

\noindent
CHAIRMAN: Dr Mendel, you’re sentenced to exile… in my Department of Extremely Safe Topics.

\noindent
DR MENDEL: What do they study?

\noindent
COMMITTEE MEMBER 1: Chairs.  (The Chairman nods approvingly)

\noindent
COMMITTEE MEMBER 2: Things with no real-world impact. (He exchanges a glance with the Chairman, maybe he is next.) (A GUARD drags DR MENDEL away.)

\noindent
DR MENDEL (shouting): What if I just explore the ethics of software licensing?
(Silence. Nervous glances.)

\noindent
CHAIRMAN (whispering): Did he say… ethics?

\noindent
(A red emergency button marked “Deploy The Clowns” is pressed. Alarms blare. The walls collapse, revealing the entire committee was inside a giant box labelled “DEPARTMENT OF NOT THINKING TOO HARD.”)
A GENERAL parachutes in, waving a giant rubber stamp.

\noindent
GENERAL: Someone thinking again? 

\noindent
COMMITTEE (pointing): Him!

(DR MENDEL now wears a dunce cap: “Academic Rebel.”)

\noindent
DR MENDEL: You can’t punish thoughts!

\noindent
GENERAL (stamps forehead): EXTREME THOUGHT CRIMINAL.
(DR MENDEL is thrown into a bottomless file drawer labelled “The Archives.” Slam.)

\noindent
DR MENDEL (muffled): But what Academic Integrity? 

\noindent
LOUDSPEAKER:  Academic Integrity has been redirected to the Department of Things We Agree With.
(A band of LOBSTERS marches in playing Rule Britania. A giraffe wanders past, reading a state-approved paper. MUSIC PLAYS. BLACKOUT.)

\endgroup
\newpage
\section*{The Archivist’s Dilemma}
\subsection*{The Setup:}
You are a researcher in a dystopian future where all scientific knowledge is controlled by The Grand Archive, a bureaucratic institution tasked with preserving “safe” information while erasing anything deemed too controversial, radical, or inconvenient.

One day, you are granted a rare one-time access to the Archive to retrieve a single research paper. However, there’s a catch:

The Archive is guarded by two Archivists—one of them always tells the truth, and the other always lies.

\subsection*{The Rules}

\begin{itemize}
\item You may ask one question to either Archivist.
\item The paper you need exists but has been censored.
\item If you retrieve it, you will never be allowed into the Archive again.
\item	If you fail, the paper will be permanently erased.
\end{itemize}

To make matters worse, the Archivists know something you don’t:
There is an exact duplicate of the research paper hidden somewhere in the Archive… but one Archivist is sworn to deny its existence.

\subsection*{The Ethical Challenges:}
\begin{enumerate}

\item Should you attempt to retrieve the censored version, knowing you may be able to reconstruct the missing parts… or risk everything searching for the duplicate?

\item If knowledge is controlled by an institution that decides what is “acceptable,” is any research ever truly recoverable?

\item Can truth exist in a system where one source is programmed to always lie?

\item Does asking the right question matter if the system itself is designed to obscure reality?
\end{enumerate}

The wrong question means the knowledge is lost forever.

The right question may only lead to a half-truth.

What do you ask?

\vfill
\section*{Diary of a Working Wanderer}

Hello Diary,

Today, I encountered something even more frustrating than trying to get a visa on arrival in Phnom Penh when the rules have mysteriously changed overnight. More maddening than negotiating a contract in Colombo last week where everyone nodded in agreement but somehow, nothing actually happened.

I tried to read a board paper…an \textit{important} one, mind you….but instead of words, I got a collection of thick, unapologetic black lines.

"\textit{Those words are forbidden}"—but, of course, they're \censor{Never gonna}. And oh boy, is it frustrating.

Now, I’ve worked around a bit to know that information flows differently everywhere. In some places, people whisper the truth like it’s an ancient spell. In others, they drown it in so much bureaucracy that by the time you reach the answer, you’ve forgotten the question. But this? This is a new one.

Someone—probably the same kind of person who thinks “transparency” means “we’re telling you \textit {just} enough to keep you quiet”— dismissed it as \textit{woke nonsense}. Because, apparently, if you don’t like the information, you can just slap a label on it and pretend it doesn’t matter. But let’s be honest: when you erase words, people don’t just stop asking questions. They start filling in the blanks. And having been to both Phnom Penh and Colombo, I can tell you — the versions people \textit{invent} are usually far wilder than the truth.

Misinformation spreads like wildfire. In fact, I once saw actual wildfire in Cambodia’s dry season—it moves fast, unpredictable, and by the time people react, it’s already out of control. Same with half-truths and convenient omissions. But redaction? That’s worse. It’s like pretending the fire doesn’t exist, even when you can smell the smoke.

I can see it now—every word blacked out, entire industries disappearing, history rewritten before our eyes. And in the background, just a little too hazy to be real, a building. Or at least, what used to be a building. The edges blur, the details vanish, and…ah yes…there’s definitely fire.

I’d ask what happened. But something tells me that’s also \censor{give you up}.

\vspace{30pt}
\newpage

\twocolumn

\section{Analysis: Critical Meaning in Six Redacted Narratives}
This analysis takes a random sample of six of the 41 stories. It delves more deeply into their critical meaning to construct a sense of what they say about creeping censorship and the limiting of academic freedoms based on a political or ideological agenda.
\begin{itemize} \item \textit{Winnie the Pooh} – When this familiar character from our childhood confronts a book with redacted content, he offers the disarmingly simple yet profound observation: \begin{quote} ``When you erase words, people don’t just forget them... they start guessing what they were... and sometimes, they guess all wrong.'' \end{quote} This line encapsulates the unsettling process by which redaction invites distortion, speculation, and unintended consequences.
\item \textit{The Forbidden Desk} – A library worker discovers a mysteriously redacted desk with sealed drawers containing blacked-out pages. As she investigates, she reflects: \begin{quote} ``People don’t forget, she realised. When you take away the truth, they don’t just let it go—they invent. And sometimes… sometimes what they create is far more dangerous than what was taken.'' \end{quote} This passage points to the double harm of redaction—erasing knowledge and inviting misinformation.
\item \textit{The Committee for Research Excellence} – A researcher faces an absurd tribunal that attempts to censor ideas before they are even spoken. The phrase "In this paper…" is treated as criminal: \begin{quote} ``First, you say, ‘In this paper…’—that’s how they get you. Then you start talking about climate change, sustainability, social justice...'' \end{quote} This satirical performance reveals the paranoia and pre-emptive control mechanisms within institutional censorship.
\item \textit{The Archivist’s Dilemma} – In a speculative future, a researcher must choose whether to access a censored document or pursue an uncensored duplicate. The story poses: \begin{quote} ``If knowledge is controlled by an institution that decides what is ‘acceptable,’ is any research ever truly recoverable?'' \end{quote} The narrative becomes a philosophical reflection on epistemic risk in bureaucratically censored systems.
\item \textit{In the Beginning There Was [REDACTED]} – A poem reimagines Biblical creation as a censored tech project: \begin{quote} ``What he forgot to do… what the client didn’t ask for… what would have saved the day… was [REDACTED].'' \end{quote} This line underscores how foundational truths can be omitted under the guise of compliance or creative constraint.
\item \textit{Diary of a Working Wanderer} – A global traveller connects the redaction of public documents with lived bureaucratic confusion: \begin{quote} ``Redaction? That’s worse. It’s like pretending the fire doesn’t exist, even when you can smell the smoke.'' \end{quote} This quote links personal frustration to global patterns of institutional opacity. \end{itemize}
\subsection*{Critical Meaning and Narrative Arc}
The six highlighted narratives collectively construct a powerful critique of information control and censorship. Their central message is that censorship doesn’t merely remove information but transforms how we understand reality, creating more dangerous conditions than the truths being suppressed. The stories show a crucial progression, revealing the machinery, impact and ultimate consequences of systematic knowledge suppression:
\begin{itemize} \item \textbf{The fragility of knowledge} – From scientific papers to children’s books to creation myths, no form of knowledge is immune from redaction (or subtle illicit change), suggesting that censorship ultimately attacks the foundations of shared reality itself.
\item \textbf{The psychological impact of information gaps} – Through characters like Winnie the Pooh and the library worker in \textit{The Forbidden Desk}, we witness the profound disorientation and frustration that results when people encounter redacted information.
\item \textbf{The dangerous act of substitution} – Every story shows how censored information doesn’t simply disappear but gets replaced—with speculation, misinformation or politically convenient alternatives that may be more destructive than the suppressed truths.
\item \textbf{Censorship as institutional policy} – From the absurdist \textit{Committee for Research Excellence} to the dystopian \textit{Archivist’s Dilemma}, the stories show how censorship becomes normalised and institutionalised through bureaucratic structures that appear at first sight to be reasonable guardians of acceptable discourse. \end{itemize}
Together, the six stories present a cohesive narrative that moves from discovery to consequences:
\begin{enumerate} \item \textbf{Discovery and confusion} – Characters first encounter redaction and censorship, experiencing bewilderment (Winnie the Pooh, the library worker, the traveller).
\item \textbf{Recognition of a system} – They come to understand censorship as deliberate and systematic rather than random or isolated (\textit{The Committee for Research Excellence}, \textit{The Archivist’s Dilemma}).
\item \textbf{Insight into the mechanisms} – The stories reveal how censorship and knowledge suppression operate through authority, bureaucracy and the manipulation of foundational narratives (\textit{A Poem}).
\item \textbf{Consequence and resistance} – Finally, they show how censorship creates information voids, sparking various forms of resistance—from the library worker’s investigation to the poem’s subversive reframing. \end{enumerate}
\subsection{Learning from our stories} 
Widening the lens to consider again the full collection \cite{guruge2025redactedBook}, the stories can now be seen to explore multiple dimensions of censorship’s impact. Our exploration ranges from dangerous information voids that breed misinformation and confusion to a sophisticated meta-awareness of narrative construction and control through power dynamics. These stories reveal how systematic erasure extends beyond words to undermine a society’s capacity to engage with critical issues. Yet, through diverse storytelling approaches, the collection itself becomes an act of creative resistance. They show us that literary innovation can be a way to circumvent and expose the machinery of censorship, to articulate what might otherwise remain unspoken.

The stories presented through this FictoLimits project were not simply fictional exercises—they were responses to an environment increasingly hostile to open inquiry. By engaging in constrained storytelling, contributors exposed the multi-dimensional nature of redaction and censorship. These stories provide insight into not only the sociopolitical implications of information control but also into how scholars and communities adapt, resist, and reimagine knowledge.

One of the most striking themes across the stories is how redaction provokes creative resistance. Contributors instinctively constructed new forms of narrative to fill voids left by erasure. For example, the \textit{Manifesto for a Fictional Movement} outlines a revolutionary vision without stating a single concrete policy, instead building solidarity around what cannot be said. This theme—of filling gaps with meaning—is echoed in \textit{The Forbidden Desk} and \textit{Those Words Are Forbidden}, where characters reconstruct history and policy from scraps and gossip. Such responses illustrate that censorship rarely silences; it shifts the terrain of expression.

Several stories also highlight the dangers of substitution and speculative reconstruction. In redacted contexts, people do not passively accept absence—they infer, reinterpret, and sometimes imagine far worse realities. This epistemic instability is depicted powerfully in stories like \textit{Those Words Are Forbidden} and \textit{Page in a Diary}, where official silence leads to widespread distrust and conspiracy:

\begin{quote}
People don’t just stop asking questions. They start filling in the blanks... the versions people invent are usually far wilder than the truth.\\
\hfill\textit{(Page in a Diary)}
\end{quote}

Importantly, redaction is never neutral. As seen in \textit{A Police Report} and \textit{Page in a Diary}, silence is used to shield power or enforce ideological uniformity. The procedural tone of these pieces mimics institutional voice while revealing the absurdity of suppressing the very data meant to inform. In \textit{Diary of a Working Wanderer}, the narrator reflects on how such practices are normalized in many countries, suggesting that resistance often takes the form of oral memory or shared inference rather than written rebuttal:

\begin{quote}
``In some places, people whisper the truth like it’s an ancient spell… But redaction? That’s worse. It’s like pretending the fire doesn’t exist, even when you can smell the smoke.''
\end{quote}

Beyond content, several stories use form and medium as sites of resistance. \textit{The Alphabet Game} and the unnamed \textit{UML/ERD Schema} pieces demonstrate how visual structure, diagram, or game mechanics can convey meaning outside traditional textual modes. These alternative formats resist algorithmic redaction while inviting new interpretive strategies:

\begin{quote}
``Trustworthy logic paths allow us to form tools that aid all, without harm or bias.''\\
\hfill\textit{(Alphabet Game)}
\end{quote}

The stories also problematize the notion of authorship and scholarly communication. In \textit{An Abstract of a Research Paper}, the structure of academic language remains while its content is entirely blacked out. This juxtaposition reveals the hollowness of bureaucratic legitimacy when stripped of critical substance:

\begin{quote}
``XXXXXXXX team-based XXXXXXX in a XXXXXXXX XXXXXX environment... Sampling of XXXXXXX XXXXXX ... benefits over traditional XXXXXXX.''
\end{quote}

Likewise, proposals to publish anonymised or redacted author lists foreground the ethics of visibility in research dissemination.

Some of the most conceptually rich stories emphasize that this methodology—fictomorphosis—is topic-agnostic. Whether dealing with domestic disagreements (\textit{Scene in a Play}), surreal limericks (\textit{Breaking Waves}), or conference absurdities (\textit{Overheard Conversation}), the method reframes how we encounter truth, omission, and inference. Genre becomes a tool for epistemic estrangement:

\begin{quote}
``If you don’t know why I’m mad, you should sleep in the guest bedroom until you figure it out.''\\
\hfill\textit{(Scene in a Play)}
\end{quote}

This curatorial and formal experimentation is itself an argument. Stories were intentionally arranged for affective and thematic resonance. Pieces like \textit{Cardboard Box Performance Thoughts} act as narrative breaths between heavier texts. This sequencing invites reflection not just on what is told, but on how reading itself becomes a resistant practice:

\begin{quote}
``Box everything up… Flatten everything.''
\end{quote}

Finally, the parallels to education, particularly computing education, are acute. In \textit{A Written Portrait} and \textit{A Dream Someone Had…}, the authors mirror the disorientation experienced by students who cannot access disciplinary language. These stories suggest that exclusionary systems—academic or political—produce similar forms of alienation, confusion, and self-silencing:

\begin{quote}
``It simply inserted a black rectangle and moved on.''\\
\hfill\textit{(Written Portrait)}
\end{quote}
\begin{quote}
``Every word that carried meaning… I just couldn’t make the sounds I needed.''\\
\hfill\textit{(A Dream Someone Had But Couldn’t Quite Explain)}
\end{quote}
The stories offer crucial lessons about the paradoxical nature of censorship. The redacted spaces breed speculation, they lead us to fill the gaps with narratives potentially more dangerous than the suppressed truths. Censorship becomes insidiously normalised through bureaucratic structures presenting themselves as ‘reasonable’ guardians of discourse. These stories highlight how censoring thoughts, research, and writing threatens the universality of knowledge, yet it ultimately defeats its own purpose by generating more radical interpretations, more radical resistance. They show us why resistance and a vigilant awareness of censoring mechanisms are essential for preserving knowledge integrity and meaningful discourse in an increasingly dystopian information landscape.

The collection provokes essential questions for our information age, challenging us to locate the point between legitimate information management and harmful censorship. It explores creative expression’s vital role as both witness to and resistance against knowledge suppression, examining how power structures determine which truths are amplified and which are erased; whose truth becomes privileged? These stories confront us with the long-term consequences of losing the language and concepts needed to address existential threats, indeed to preserve a world view. Of course, such marginalisation is itself not new but a hallmark familiar to authoritarianism and colonialism. How then can we preserve the integrity of knowledge when institutional forces systematically devalue, discredit and delete certain ideas from public discourse? Such questions become increasingly urgent as information control becomes more sophisticated, more pervasive, and words harder to legitimate.
\subsection{Learning from our process}

Fictomorphosis demonstrates that creative story transformation allows researchers to adapt, counter organised ignorance, and maintain critical inquiry in adverse conditions. 
The purpose of employing Fictomorphosis is to provide computing researchers with a means to persist in addressing crucial, socially significant topics, whether their work is disregarded as "woke," underfunded, or otherwise marginalised. Fictomorphosis enables researchers to initiate dialogues in innovative, non-confrontational manners that circumvent ideological opposition. It allows for a more ethical and safer engagement with delicate themes, redirecting the emphasis from individual culpability to the identification and resolution of overarching structural concerns. By converting life experiences into fictionalised tales, researchers can safeguard anonymity, uphold critical inquiry, and sustain the integrity of their work, even in hostile or censored contexts.

The authors engaged in deep discussion about the implications of writing this piece. Should we redact our names? Would publishing it expose us to further challenges? In the end, we chose to use our names — what’s the worst that could happen? Yet, in doing so, we acknowledge that this decision reflects our own positions of relative power and security. We must ask: does the accountability for the transgressive nature of the original subject matter simply shift from those subjects onto us, the authors of these fictomorphosised versions? Is there an implicit presumption that the author, in creatively re-narrating these stories, occupies a position of greater social power — able to "get away with it" in ways that the original subject could not?

The stories presented here are emotionally and ethically true, though creatively re-narrated. This strategy seeks to protect individuals while exposing the emotional truth of lived experiences. Yet this form of redaction risks creating the impression that there is nothing to see — and, therefore, no reason to resist. Fictomorphosis attempts to recover the emotional force of these suppressed stories, albeit imperfectly. This act of re-telling is not intended to detract from the original lived experiences or to redirect accountability. Rather, it seeks to honour what cannot be safely spoken and to preserve emotional and political truth in an ethically safe and imaginative form.

This narrative form, while derived in part from lived emotional truths, may also reflect deeper self-identity constructions, such as our own position as researchers navigating institutional hostility. We recognise this tension and acknowledge that even ethically motivated storytelling can reproduce the dynamics it seeks to criticise. Rather than rejecting this tendency, Fictomorphosis encourages a diffractive engagement with it, allowing space for both resistance and complicity, self and system, without assuming clarity or purity. 

According to Le Guin's \emph{The Carrier Bag Theory of Fiction} \cite{leguin2019carrierbag}, instead of glorifying the spear-wielding hero who kills things, stories can be about the quieter, more sustained acts of care, survival, and interdependence — the carrier bag that holds life's messy contents. It challenges the dominant (often patriarchal) narrative arc, which focusses on conflict, conquest, and heroic individualism.

Ultimately, Fictomorphosis empowers researchers to counteract suppression while persistently championing sustainability, justice, and ethical computing practices. Our analysis offers pragmatic insights for addressing censorship, safeguarding knowledge integrity, and maintaining an environment conducive to sustainability and justice-focused computing research.

\section{Conclusion}
Perhaps the most profound insight across these narratives is that censorship is ultimately self-defeating. By attempting to control discourse through redaction, authorities paradoxically create the conditions for more radical and potentially destabilising interpretations. As Pooh observes, when you erase words, “people don’t just forget them... they start guessing.”
This collection acts as both warning and resistance—documenting the mechanisms of censorship while simultaneously showing us creative ways to work around them, resist them and to expose their internal contradictions.

Fictomorphosis is more than a narrative technique; it is an emancipatory research methodology that enables scholars to:
\begin{itemize}
\item Engage with contentious, ethically restricted, or legally constrained topics
\item Transform lived experiences into creative, ethically responsible narratives
\item Challenge dominant discourses and power structures through storytelling
\item Provide fresh methodological tools for Professional Practice researchers navigating NDAs, workplace politics, and marginalised perspectives
\end{itemize}

By reframing lived experiences through creative genres, Fictomorphosis opens new pathways for knowledge production, ensuring that critical, untold stories are explored and shared without harm. In a world where taboo topics in computing, technology, and society remain contested, this approach offers a powerful, ethical, and transformative research tool.  To read such stories, with their partial truths, gaps and unfinished endings, is to participate in the work of becoming-with others in ongoing struggle.

Beyond creative adaptation, researchers whose work is labelled as “woke” or politically sensitive must also seek collective strategies for resilience. While methods like Fictomorphosis offer one pathway, those of us whose work remains institutionally supported must double down—continuing critical research, actively collaborating with marginalised colleagues, and finding alternative venues to sustain vital conversations. Building networks of solidarity, co-authoring across perceived boundaries, and amplifying suppressed areas of inquiry are essential. Constraint demands not retreat, but ingenuity, mutual reinforcement, and strategic persistence.

\balance

\bibliographystyle{ACM-Reference-Format}
\bibliography{Guruge25_Fictolimits}

\end{document}